\documentclass[traditabstract]{aa} 
 
\usepackage{amsmath, graphicx, times, txfonts, float, wasysym, color}

\newcommand{\less}{\raisebox{-1.1mm}{$\stackrel{<}{\sim}$}}

\newcommand{\msolyr}{{$M_{\odot}$}\,yr$^{-1}$} 
\newcommand{\mdot}{$\dot{M}$}
\newcommand{\lsol}{\mbox{$L_{\odot}$}} 
 
\newcommand{\kks}{K kms$^{-1}$} 
\newcommand{\ks}{km s$^{-1}$} 
 
\newcommand{\mum}{$\mu$m}

\begin{document}

\title{
The photodissociation of CO in circumstellar envelopes\thanks{
Table~\ref{Tab-Res} is only available in electronic form at the CDS via 
anonymous ftp to cdsarc.u-strasbg.fr (130.79.128.5) or via 
http://cdsweb.u-strasbg.fr/cgi-bin/qcat?J/A+A/. 
} 
}  
 
\author{ 
M.~A.~T.~Groenewegen 
}

\institute{ 
Koninklijke Sterrenwacht van Belgi\"e, Ringlaan 3, B--1180 Brussels, Belgium \\ \email{martin.groenewegen@oma.be}
} 
 
\date{received: 2017, accepted: 2017} 
 
\offprints{Martin Groenewegen} 
 
 
\abstract {
Carbon monoxide is the most abundant molecule after H$_2$ and is important for chemistry in circumstellar 
envelopes around late-type stars. The size of the envelope is important when modelling  
low-J transition lines and deriving mass-loss rates from such lines.
Now that ALMA is coming to full power the extent of the CO emitting region can be measured directly for 
nearby asymptotic giant branch (AGB) stars.
In parallel, it has become obvious in the past few years that the strength of the interstellar radiation field (ISRF) 
can have a significant impact on the interpretation of the emission lines.

In this paper an update and extension of the classical Mamon et al. (1988; ApJ 328, 797) paper is presented; 
these authors provided the CO abundance profile, described by two parameters, as a function of mass-loss rate and 
expansion velocity.
Following recent work an improved numerical method and updated H$_2$ and CO shielding functions are used 
and a larger grid is calculated that covers more parameter space, including the strength of the ISRF.
The effect of changing the photodissociation radius on the low-J CO line intensities is illustrated in two cases.
}

\keywords{Astrochemistry -- Stars: AGB and post-AGB -- Stars: winds, outflows -- Radio lines: stars} 

\maketitle

\section{Introduction} 

Nearly three decades ago Mamon et al. (1988; hereafter MGH) published a paper with the same title as the present paper.
These authors calculated the unshielded CO photodissociation rate ($I_0 = 2.0 \cdot 10^{-10} s^{-1}$), and found it to 
be a factor 10 larger than earlier studies used at the time ($2 \cdot 10^{-11} s^{-1}$; Federman et al. 1980).
Mamon et al. used a schematic one-dimensional model to calculate the CO distribution in the envelopes around evolved stars.
This distribution was approximated by a two-parameter model (see Eq.~\ref{Eq-M}), and the parameters were
given in tabular form as a function of a range of mass-loss rates and three expansion velocities.
Very quickly, several analytical fit formulae to this tabular data were presented 
(Planesas et al. 1990, Kwan \& Webster 1993, Stanek et al. 1995).

The original MGH results and the analytical formulae are still widely in use today to estimate 
qualitatively the size of the CO shell and whether such a shell would be resolved by a 
single-dish radio telescope. The MGH results are also used in molecular radiative transfer (RT) codes to fix the CO abundance profile 
as a function of distance to the central star for a given (assumed) mass-loss rate and expansion velocity.

Almost 30 years later it seems timely to revisit the problem. 
The unshielded CO photodissociation rate has been revised upwards by about 30\% (Visser et al. 2009) and detailed
shielding functions are available (Visser et al. 2009). The MGH study used some analytical approximations for 
the two-dimensional problem in the calculation of the shielded photodissociation rate as a function of distance 
to the star. These approximations are not needed and the two-dimensional problem can be solved numerically (Li et al. 2014).
On the observational side, the ALMA interferometer is reaching its full potential, and CO envelopes can be studied 
in much more detail. The spatial extent of the molecular shells of nearby evolved stars can be determined 
accurately, and therefore detailed theoretical predictions of the size of the CO shell seem timely.

Li et al. (2014, 2016) recently incorporated the shielding functions by Visser et al (2009) and the N$_2$
shielding function (Li et al. 2013) together with a numerical integration scheme (dubbed "the spherically symmetric model") into a
chemical network code to calculate abundance profiles for various molecules in the envelopes around IK Tau and CW Leo.
This model and the CO shielding functions are at the basis of the present work.

It has also become clear in recent years that the strength of the radiation field plays an important role.
The MGH work used the interstellar radiation field (ISRF) by Jura (1974), calculated for the solar vicinity by
explicitly considering the Sun and 14 stars earlier than B5 listed in the {\it Catalogue of Bright Stars}.
At the five wavelengths between 930 and 1125 \AA\ listed by Jura the ISRF by Draine (1978) agrees to within 5\%. 
Draine also provides estimates for the ISRF above the Galactic plane and, for a range of parameters, typical strengths 
are between 0.2 and 0.8 times the value in the Galactic plane.
 
Doty \& Leung (1998) considered the effect of the strength of the ISRF in their model of the envelope 
around IRC~+10\degr~216 (CW Leo). Their results are a bit difficult to estimate (see one of the tiny panels in their Fig.~8) 
but suggest a decrease in the photodissociation radius by a factor of $\sim4$ when increasing the strength 
of the ISRF by a factor of 9.

McDonald et al. (2015) presented the results of ALMA observations of evolved stars in the globular cluster 47 Tucanae.
Four of the brightest AGB stars with estimated mass-loss rates of a few  $10^{-7}$ \msolyr\ based on 
their infrared emission were observed in the CO J= 2-1 line. None of these stars were detected although the observations
were sensitive enough to detect line strengths predicted by standard formulae (Olofsson 2008, and 
Ramstedt et al. 2008 for details) based on observations of a range of Galactic AGB stars.
McDonald et al. (also see McDonald \& Zijlstra 2015) estimated the ISRF to be typically 50 times stronger 
than the Galactic ISRF, depending om the location of the star in the cluster and other factors, 
but stronger than a factor of 2.5 with a probability of at least 85\% for all four objects. McDonald et al. also estimated 
that the CO envelopes were truncated at a few hundred stellar radii and that the line intensities were 
about two orders of magnitude below their current detection limits. These authors also found the radiation field to be harder
than the Draine field.
Zhukovska et al. (2015) showed the importance of the ISRF in the photodissociation of SiO in clusters.
These authors found that AGB stars in clusters experience an ultraviolet (UV) field typically 10-100 times the ISRF.
Groenewegen et al. (2016) discussed the non-detection of CO J= 2-1 in one and the marginal detection in 
another OH/IR star in the LMC.
There are several possible reasons for this, but one is the stronger radiation field in the LMC diffuse medium 
by a factor of $\sim 5$ (Paradis et al. 2009) compared to the Galactic Draine field.

In Sect.~2 the physical problem and the mathematical and numerical solution to this problem are presented.
The calculations are presented in Sect.~3 and discussed in Sect.~4. Section~5 concludes this paper.

\section{Equations and solutions} 

Consider a spherically symmetric homogeneous outflow with expansion velocity $V$.
If photodissociation is the only destruction mechanism and in the absence of molecular formation processes, 
the number density, $n$, of CO is given by (Jura \& Morris 1981)
\begin{equation}
\frac{1}{r^2} \frac{\partial}{\partial r} (r^2 \; n \; {\rm V}) = -I(r) \; n, 
\end{equation}
with $I$ the photodissociation rate.
Using $(r^2 \; n \; V)$ as variable, the solution of the CO abundance relative to H$_2$, and relative to the value 
at the inner radius, for constant velocity is
\begin{equation}
x (r_{\rm i}) = x (r_{\rm i-1}) \, \exp \left( - \frac{1}{\rm V} \; \int_{r_{\rm i-1}}^{r_{\rm i}} I(r^\prime) \, d r^\prime \right),
\label{Eq-Sol}
\end{equation}
with the boundary condition that at the inner radius $x (r_{\rm 1}) = 1$.
In the case of unshielded radiation, $I (r) = I^0$, and the solution becomes 
\begin{equation}
x (r) = x (r_0) \exp \left( - \frac{I^0}{\rm V} \; r \right).
\label{Eq-Uns}
\end{equation}

The shielded dissociation rate at a point in the envelope depends on the conditions in the entire shell
\begin{equation}
I (r) = \frac{1}{2} \int_{0}^{\pi} k (r, \theta) \sin \theta \; d \theta,
\end{equation}
where $k (r, \theta)$ is the dissociation rate at a radius $r$ from the central star by 
interstellar photons along a ray making an angle $\theta$ (Li et al. 2014, see Figure~\ref{Fig-Sit}).

The integration over $\theta$ effectively goes from 0 to  $(\pi - \beta),$ where 
\begin{equation}
\sin \beta = \frac{R_{\star}}{r_{\rm i}}
\end{equation}
indicates the angle subtended by the central star (or the inner radius of the envelope) from point $r_{\rm i}$.
The UV radiation field from the cool AGB can be neglected. The numerical code could be adapted however 
to include the radiation from a hotter central object, simulating either a corona or chromosphere around the central star 
or a close binary component. This would make the problem time dependent however.

The dissociation rate can be written as (Li et al. 2014)
\begin{equation}
k (r, \theta) = \chi \; I_0 \; \Theta_{\rm dust}(r, \theta) \; \Theta_{\rm H_2, CO} (r, \theta),
\label{Eq-k}
\end{equation}
where $I_0 = 2.6 \cdot 10^{-10} s^{-1}$ (Visser et al. 2009) is the unshielded photodissociation rate, 
$\chi$ is a scaling factor indicating the strength of the ISRF relative to the Draine (1978) field 
adopted in Visser et al., and $\Theta_{\rm dust}= \exp \, ( -\tau_{\rm dust} )$ the shielding by dust 
and $\Theta_{\rm H_2, CO}$ the $^{12}$CO self-shielding and shielding by H$_2$ and the CO isotopologues (Visser et al. 2009).
The tabulated shielding function in Visser et al. depend on the CO and H$_2$ column density, 
and the excitation temperature\footnote{The files are available at 

http://home.strw.leidenuniv.nl/$\sim$ewine/photo/

index.php?file=CO\_photodissociation.php. 
The files calculated for a Doppler with of 0.3 \ks\ for CO, and a $^{12}$CO/$^{13}$CO ratio of 69 have been used.}.

This immediately implies that the solution to Eq.~2 must be obtained by iteration, as the CO photodissociation 
rate in a point in the  circumstellar envelope (CSE) depends on the CO column density along all rays (see Figure~\ref{Fig-Sit}).
The solution proceeds as follows (following Li et al.):
\begin{itemize}
\item Assume a CO profile in the envelope.
The parametrisation introduced by MGH is used, i.e.
\begin{equation}
x (r) = x_0 \exp\left( - \ln(2) \left( \frac{r}{R_{\frac{1}{2}}} \right) ^{\alpha} \right).
\label{Eq-M}
\end{equation}

\item
In point $r_{\rm i}$ determine the column densities of CO and H$_2$ (see Appendix~\ref{App-Num}) and 
the dust optical depth along all angles, and perform the integration over $\theta$. 

\item 
With $I(r_{\rm i})$ determined, 
calculate $x(r_{\rm i})$ from Eq.~\ref{Eq-Sol}, and proceed to the next radial point.

\item
After the full profile is determined, fit Eq.~\ref{Eq-M} to it, and determine new estimates for 
$R_{\frac{1}{2}}$ and $\alpha$. Iterate until these two parameters no longer change.

\end{itemize}

Input parameters to the model are 
the total mass-loss rate in \msolyr, 
(constant) gas expansion velocity in the CSE (V), 
number ratio of helium to hydrogen ($f_{\rm He}$), 
CO abundance relative to H$_2$ at the inner radius ($f_{\rm CO}$), 
and properties of the dust.
Although the results presented here are for parameters typical of AGB star winds, 
the code is set up in a flexible way that allows the modelling of individual objects. 
To facilitate that, the dust-to-gas ratio ($\Psi$), grain specific density ($\rho_{\rm g}$), 
grain size ($a_{\rm g}$), and dust extinction coefficient at 1000 \AA\ ($Q_{\rm e}$) are the input
parameters, although the results only depend on the combination 
($\frac{Q_{\rm e} \, \Psi}{\rho_{\rm g} \, a_{\rm g}}$) (Eq.~\ref{Eq-dust}).
In other codes the prescription of the dust extinction is more general, for example Li et al. (2014) 
essentially used the H$_2$ column density and a standard conversion factor (Bohlin et al. 1978, Rachford et al. 2009) 
to determine the extinction. Mamon et al. and others (Nejad \& Millar 1987, Doty \& Leung 1998)
typically assumed an optical depth in the UV at a given inner radius (Eq.~\ref{Eq-tau}), which 
scales with the assumed mass-loss rate and velocity (Eq.~\ref{Eq-dust}). 
Doty \& Leung (1998) found that for parameters typical for CW Leo a change in dust optical depth by a factor
of five has a very small effect on the CO profile.

Less important parameters are the inner radius of the shell, where the calculation starts---a few stellar radii, where 
the stellar radius is determined from the stellar luminosity and the effective temperature of the central star, 
which are input parameters--and the outer radius, which is arbitrarily set to 15~$R_{\frac{1}{2}}$. 
In the limiting case of $\alpha \rightarrow 1$ for very low mass-loss rates the relative abundance has 
dropped to $\sim 3 \cdot 10^{-5}$ at that distance.

Finally, the CO excitation temperature profile needs to be provided.
The MGH study took a gas kinetic temperature profile that is derived for CW Leo and assume this profile is 
valid in all their calculations. These authors assumed that the excitation 
temperature equals the gas temperature, and for parameters typical for CW Leo test the case that the 
excitation temperature is half the gas temperature finding that $R_{\frac{1}{2}}$ is increased by $\sim20\%$.
Doty \& Leung (1998) found that for parameters typical for CW Leo, a factor of two change in gas temperature 
has a small effect on the CO abundance profile.
Li et al. (2014), in their model for CW Leo, took the excitation temperature to be constant in the CSE and equal to 
5~K, as they found that CO is subthermally excited at the low densities in the outer parts of the CSE.
The CO models for OH 32.8$-$0.3 and OH 44.8$-$2.3 (Groenewegen 1994b), the halo carbon star 
IRAS 12560+1657 (Groenewegen et al. 1997), and CW Leo (Groenewegen et al. 1998) indeed show this to be the case. 
In these models, the excitation temperatures of the J= 1-0 to 4-3 transitions range from 3 to 12~K in the regions 
of the wind where the CO abundance has dropped to $\sim 0.8$ to $\sim 0.3$ of its initial value, and the excitation 
temperatures are always lower, and sometimes lower by a factor of a few, than the gas temperature at these radii.
Additional details on the calculations are provided in Appendix~\ref{App-Num}.

\section{Calculations} 

A large model grid was calculated using the following parameters:

\begin{itemize}

\item 
Seventeen mass-loss rates (MLRs) between $1 \cdot 10^{-13}$ and $2 \cdot 10^{-4}$ \msolyr.
The lowest MLR could never be measured in practice but allows a numerical check of the results in the case of
unshielded radiation.

\item
Expansion velocities of 3.0, 7.5, 15 and 30 \ks.

\item 
Nine CO abundances between $0.1 \cdot 10^{-4}$ and $12 \cdot 10^{-4}$.

\item
Nine different strengths of the ISRF, $\chi=$ 0.01, 0.1, 0.2, 0.5, 1, 2, 5, 10, and 100.

\item
Four constant excitation temperatures of 5, 20, 50, and 100~K.

\end{itemize}

\noindent
The dust properties are assumed to be 
$\Psi= 0.005$,
$Q_{\rm e}= 4.5$,
$a_{\rm g}= 0.15$~\mum,
$\rho_{\rm g}= 1.8$ g cm$^{-3}$,
which are typical for the dust properties around Galactic AGB stars. 
This corresponds to an UV dust optical depth of about 21 at $1 \cdot 10^{15}$ cm.
The inner radius is assumed to be $R_{\rm in}= 4~R_{\star}$, where the stellar radius is calculated from
$L= 8000$~\lsol\ and $T_{\rm eff}= 2600$~K, resulting in an inner radius of $1.2 \cdot 10^{14}$ cm.       
The helium abundance is assumed to be $f_{\rm He}= 0.08$.
The resulting values of $R_{\frac{1}{2}}$ and $\alpha$ are given in Table~\ref{Tab-Res}, which is available 
in its entirety at the CDS. 
Not all possible combinations are listed in the Table. For the lowest MLRs and the largest ISRF strengths the calculated
dissociation radius is smaller than the adopted inner radius.
In Table~\ref{Tab-Res} selected models are listed to illustrate the general behaviour of the results.
The models are for standard values of $f_{\rm CO}= 8 \cdot 10^{-4}$, expansion velocity 15~\ks, standard ISRF, 
and excitation temperature of 5~K. The table also lists the results when changing these parameters.
The values found by MGH are listed for comparison for the various MLRs. A value of $1 \cdot 10^{-5}$ \msolyr\ 
is the standard value considered in the discussion below. The standard value for MLR, CO abundance and expansion velocity
adopted here are the same as in the standard model in MGH.

For the standard model, Table~\ref{Tab-Res} lists the result when changing the dust properties by a factor of 3 
(an effect of $^{+16}_{-7}\%$ on $R_{\frac{1}{2}}$) and using CO shielding functions for a different assumed 
Doppler width or isotopic abundance ratio (see Visser et al.), with an  $\sim 1\%$ effect.

\begin{table*}

\caption{CO line profile parameters $R_{\frac{1}{2}}$ and $\alpha$ as a function of input parameters. }
  \begin{tabular}{ccccccccclcccccc}
  \hline
\mdot     & f$_{\rm CO}$ & V     & $\chi$ &  $T_{\rm ex}$ &  $R_{\frac{1}{2}}$ & $\alpha$ &  $R_{\frac{1}{2}}$ & $\alpha$       & Remark \\
(\msolyr) &            & (\ks) &        &     (K)     &   ($10^{15}$ cm) &           &    ($10^{15}$ cm) &                &   \\
          &            &       &        &             & \multicolumn{2}{c}{This work} & \multicolumn{2}{c}{MGH} &  \\
\hline
2 (-4) &  8 (-4)   & 15  & 1 & 5 & 1280    & 3.64 &      &     \\
1 (-4) &  8 (-4)   & 15  & 1 & 5 &  866.1  & 3.37 & 1070 & 3.39 \\
5 (-5) &  8 (-4)   & 15  & 1 & 5 &  597.2  & 3.15 &  667 & 3.20 \\
2 (-5) &  8 (-4)   & 15  & 1 & 5 &  375.1  & 2.91 &  365 & 2.96 \\
1 (-5) &  8 (-4)   & 15  & 1 & 5 &  268.6  & 2.77 &  235 & 2.79 & standard model \\

1 (-5) &  10 (-4)  & 15  & 1 & 5 &  288.6  & 2.73 &   &  \\
1 (-5) &  8 (-4)   & 30  & 1 & 5 &  286.2  & 2.58 &   &  \\
1 (-5) &  8 (-4)   & 15  & 0.1 & 5 &  967.  & 2.47 &  &  \\
1 (-5) &  8 (-4)   & 15  & 10  & 5 &   86.6 & 3.37 &  &  \\
1 (-5) &  8 (-4)   & 15  & 1  & 50 &  253.5 & 3.64 &  &  \\

1 (-5) &  8 (-4)   & 15  & 0.77 & - &  287.7  & 3.37 & 235 & 2.79 & Simulation of MGH, see Sect.~\ref{S-MHG} \\ 

1 (-5) &  8 (-4)   & 15  & 1 & 5 &  311.8  & 3.22 &  &  & $(\frac{Q_{\rm e} \, \Psi}{\rho_{\rm g} \, a_{\rm g}}) \cdot 3$ \\
1 (-5) &  8 (-4)   & 15  & 1 & 5 &  250.7  & 2.57 &  &  & $(\frac{Q_{\rm e} \, \Psi}{\rho_{\rm g} \, a_{\rm g}}) /     3$ \\

1 (-5) &  8 (-4)   & 15  & 1 & 5 &  271.1  & 2.98 &  &  & CO Doppler width of 3~\ks \\
1 (-5) &  8 (-4)   & 15  & 1 & 5 &  269.7  & 2.78 &  &  & $^{12}$CO/$^{13}$CO= 35     \\


5 (-6) &  8 (-4)   & 15  & 1 & 5 &  194.2  & 2.67 &  154   & 2.61 \\
2 (-6) &  8 (-4)   & 15  & 1 & 5 &  129.9  & 2.55 &   88.8 & 2.39 \\
1 (-6) &  8 (-4)   & 15  & 1 & 5 &   96.72 & 2.47 &   59.5 & 2.24 \\
5 (-7) &  8 (-4)   & 15  & 1 & 5 &   72.77 & 2.38 &   40.5 & 2.09 \\
2 (-7) &  8 (-4)   & 15  & 1 & 5 &   50.81 & 2.25 &   25.4 & 1.89 \\
1 (-7) &  8 (-4)   & 15  & 1 & 5 &   39.26 & 2.12 &   18.5 & 1.74 \\
5 (-8) &  8 (-4)   & 15  & 1 & 5 &   30.74 & 2.00 &   14.0 & 1.60 \\
2 (-8) &  8 (-4)   & 15  & 1 & 5 &   22.75 & 1.832 &  10.5 & 1.46 \\
1 (-8) &  8 (-4)   & 15  & 1 & 5 &   18.46 & 1.715 &   9.01 & 1.39 \\

1 (-9) &  8 (-4)   & 15  & 1 & 5 &   10.38 & 1.439 & & \\
1 (-10) &  8 (-4)  & 15  & 1 & 5 &    6.85 & 1.291 & & \\
1 (-11) &  8 (-4)  & 15  & 1 & 5 &    5.26 & 1.160 & & \\
1 (-13) &  8 (-4)  & 15  & 1 & 5 &    4.17 & 1.058 & & \\ 
\\
1 (-15) &  8 (-4)  & 15  & 1 & 5 &    4.144 & 1.052  & & & $L=8000$ \lsol;  $R_{\rm in}= 4.0~R_\star$ \\ 
1 (-15) &  8 (-4)  & 15  & 1 & 5 &    4.045 & 1.0167 & & & $L=800$ \lsol;  $R_{\rm in}= 4.0~R_\star$\\ 
1 (-15) &  8 (-4)  & 15  & 1 & 5 &    4.014 & 1.0054 & & & $L=80$ \lsol;  $R_{\rm in}= 4.0~R_\star$\\ 
1 (-15) &  8 (-4)  & 15  & 1 & 5 &    4.004 & 1.0018 & & & $L=8$ \lsol;  $R_{\rm in}= 4.0~R_\star$\\ 
1 (-15) &  8 (-4)  & 15  & 1 & 5 &    4.001 & 1.0007 & & & $L=8$ \lsol;  $R_{\rm in}= 1.1~R_\star$\\ 

\hline
\end{tabular}

\label{Tab-Res}
\end{table*}

\section{Discussion} 

\subsection{No mass loss}
\label{Sect-NML}

In the limit of very low mass-loss rates, that is essentially unshielded radiation, 
Eqs.~\ref{Eq-Uns} and \ref{Eq-M} show that one expects 
$\alpha \rightarrow 1$ and $R_{\frac{1}{2}} = \ln(2) \; V/ (\chi \, I_0)$.
The typical radius for the abundance to change in the unshielded Draine ISRF is therefore 
$4.0 \cdot 10^{15}$ cm in an envelope with 15 \ks\ expansion velocity.

Inspection of Table~\ref{Tab-Res} shows that $\alpha$ indeed is close to unity in the models with the lowest 
mass-loss rates, but that $R_{\frac{1}{2}}$ is larger than the expected values. This is the effect of 
the finite $R_{\star}$ and $R_{\rm in}$ adopted in the models. It implies that some of the ISRF is blocked by the central star
and that the photodissociation radius is therefore slightly larger.

Table~\ref{Tab-Res} includes a model with an even lower MLR ($1 \cdot 10^{-15}$ \msolyr), models
were the stellar radius is systematically decreased (by lowering the stellar luminosity), and a model
in which  R$_{\rm in}$ is decreased from 4 to 1.1~R$_{\star}$. The half-radius and $\alpha$ indeed converge 
to the expected values in the limiting case of unshielded radiation.

\subsection{Comparing to MGH}
\label{S-MHG} 

Table~\ref{Tab-Res} lists the results found here and those listed in MGH. The behaviour is complicated. 
For the lowest MLRs the value for $R_{\frac{1}{2}}$ found here is typically a factor of about 2 larger, 
but the difference becomes smaller with increasing MLR and the values are about equal near $2 \cdot 10^{-5}$ \msolyr.

The main difference in the two approaches is in the way the shielding is calculated, but there are other differences.
Therefore an attempt was made to more closely follow the assumptions made in MGH.

Mamon et al. assumed that the CO excitation temperature follows the gas kinetic temperature, and their temperature law was adopted here.
The default dust properties were changed to give the dust optical depth at 1000~\AA\ adopted by MGH.
The inner radius is set at $10^{16}$ cm and the central star is assumed to be a point source.
The ISRF adopted in MGH is different from adopted here; the field as determined by Jura et al. (1974) 
implies $\chi= 0.77$ with respect to the Draine field.

In our model, the resulting value for $R_{\frac{1}{2}}$ is larger than in the standard model by about 7\% and is 22\% larger than MGH.
The difference between the use of "the spherically symmetric model" compared to the one-dimensional approach in MGH is difficult 
to quantify (but see Li et al. 2014). This implies that the shielding in Visser et al., which included updated molecular data 
compared to MGH, is more effective than in the older approach.

\begin{figure}

\centering
\includegraphics[width=0.99\hsize]{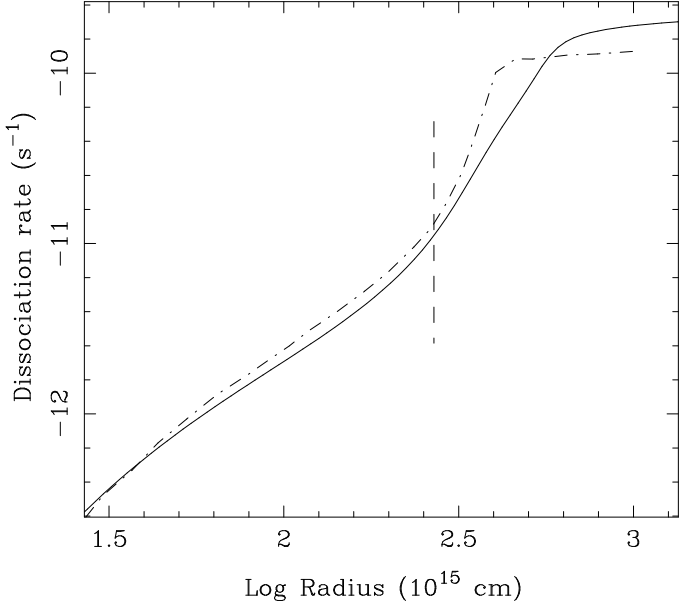}

\caption[]{ 
Solid lines indicates the CO photodissociation rate as a function of radial distance for the standard model.
The dashed line indicates the value of $R_{\frac{1}{2}}$ for that model. 
The dot-dashed line indicates the rate in the standard model of MGH (which has similar parameters).
} 

\label{Fig-Disso} 
\end{figure}

\begin{figure} 

\centering
\includegraphics[width=0.99\hsize]{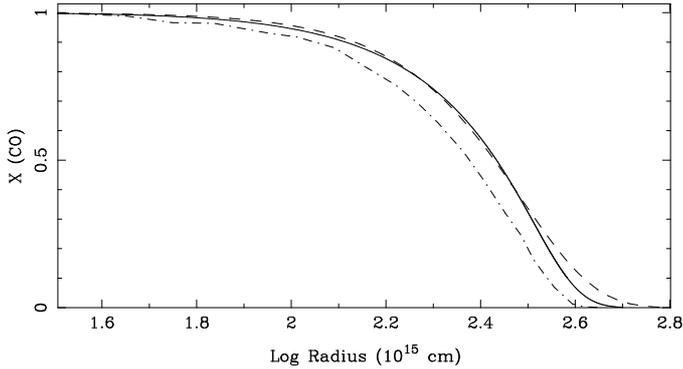}

\caption[]{ 
Solid line indicates the relative CO profile for the standard model, while 
the dashed line gives the profile for the fitted values of $R_{\frac{1}{2}}$ and $\alpha$.
The dash-dotted lines gives the profile for the similar model in MGH.
} 

\label{Fig-Prof} 
\end{figure}

\begin{figure} 

\centering
\includegraphics[width=0.99\hsize]{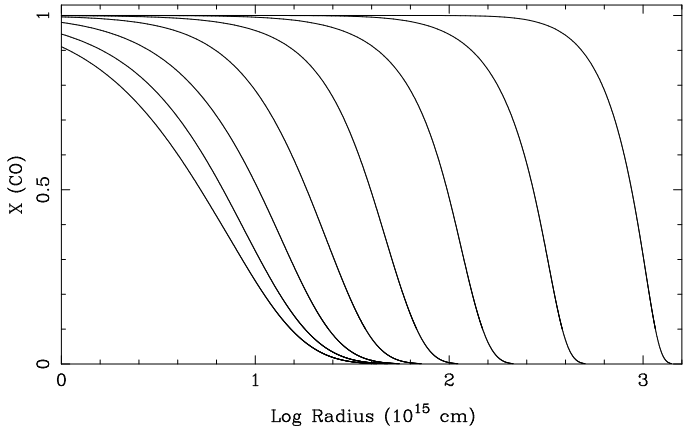}

\caption[]{ 
CO abundance profiles for MLRs between $10^{-11}$ and $10^{-4}$ \msolyr\ in steps of a decade.
} 

\label{Fig-ProfAll} 
\end{figure}

Figure~\ref{Fig-Disso} compares the dissociation rate of the standard model with MGH (their Figure~1). 
The rate is indeed larger, leading to a smaller photodissociation radius.
Figure~\ref{Fig-Prof} shows the relative CO abundance in the envelope  for the standard model and compares this model to
the MGH model (see their Figure~2), while Fig.~\ref{Fig-ProfAll} shows the profile for several MLRs (cf. Figure~3 in MGH).

\subsection{Fitting formula}

Soon after the publication of MGH, several convenient fitting formulae were published that allowed the community to estimate
$R_{\frac{1}{2}}$ and $\alpha$ for a large parameter space.
The formula in Planesas et al. (1990) is
\begin{equation}
R_{\frac{1}{2}}  =  62 \; \left(\frac{\dot{M}}{10^{-6}}\right)^{0.58} \; \left(\frac{V}{15}\right)^{-0.4} \; \left(\frac{f_{\rm CO}}{4 \cdot10^{-4}}\right)^{0.5}; 
\end{equation}
Kwan \& Webster (1993) derived
\begin{equation}
\alpha  =  2.27 \; \left(\frac{\dot{M}}{10^{-6}} \cdot \frac{15}{V}\right)^{0.09}, 
\end{equation}
and (for the value of $f_{\rm CO}= 8 \cdot 10^{-4}$ assumed by MGH)
\begin{equation}
R_{\frac{1}{2}}  =  56 \; \left(\frac{\dot{M}}{10^{-6}}  \cdot \frac{15}{V}\right)^{0.63}.
\end{equation}
The most elaborate approximation formula is that by Stanek et al. (1995), who considered the analytical value in the 
limit of small mass loss rates,
\begin{equation}
R_{\frac{1}{2}}  =  54 \; \left(\frac{\dot{M}}{10^{-6}}\right)^{0.65} \; \left(\frac{V}{15}\right)^{-0.55} \; \left(\frac{f_{\rm CO}}{8 \cdot 10^{-4}}\right)^{0.55} 
 + 7.5 \; \left(\frac{\rm V}{15}\right) 
\end{equation}

This functional form is used here as well and the fit is shown in Eq.~\ref{Eq-Rh}.
This fit was derived as follows. The second term on the right-hand side is the value in the limit of unshielded radiation and has an 
analytic result, where a correction was applied because of the finite size of the central star; see Sect.~\ref{Sect-NML} 
($f_{\rm sc}= 4.144/4 = 1.036$; see Tab.\ref{Tab-Res}).
This term was subtracted from the calculated value of $R_{\frac{1}{2}}$ and then a multi-dimensional linear fit in logarithmic space 
was made using the {\it SVDFIT} routine available in {\it Numerical Recipes} (Press et al. 1992).
The fit was restricted to models with MLRs larger than $10^{-8}$ \msolyr\ and 0.1 $< \chi <$ 10.
In addition models with velocities of 3.0 \ks\ were excluded for MLRs larger than $5 \cdot 10^{-5}$ \msolyr; these models showed 
larger residuals and in fact such combinations of parameters are not observed.
Similarly,  Eq.~\ref{Eq-Alf} shows the fit for $\alpha$, i.e.

\begin{equation}
\begin{aligned}
R_{\frac{1}{2}} & = & 42 \; \left(\frac{\dot{M}}{10^{-6}}\right)^{0.57}   \; \left(\frac{V}{15}\right)^{-0.14} \; 
                         \left(\frac{f_{\rm CO}}{10^{-4}}\right)^{0.25} \; \chi^{-0.41}  \; \left(\frac{T_{\rm ex}}{10}\right)^{+0.02}    \\
             &  & + f_{\rm sc} \; 4.00 \; \left(\frac{\rm V}{15}\right) \; \left(\frac{2.6 \cdot 10^{-10}}{k_0}\right) \; \frac{1}{\chi}
\end{aligned}
\label{Eq-Rh}
\end{equation}
\begin{equation}
\alpha = 1 +  1.45 \; \left(\frac{\dot{M}}{10^{-6}}\right)^{0.16}    \; \left(\frac{\rm V}{15}\right)^{-0.31} \; 
                      \left(\frac{f_{\rm CO}}{10^{-4}}\right)^{-0.01}  \; \chi^{0.16} \; \left(\frac{T_{\rm ex}}{10}\right)^{+0.08}
\label{Eq-Alf}
\end{equation}

Although the fitting formula may be convenient, it is definitely recommended to interpolate in the Tab.~\ref{Tab-Res} directly.
The uncertainty in the power law fit is important; i.e. 
to have the fitting routine return a reduced $\chi^2$ of unity, an uncertainty of about 24\% in $R_{\frac{1}{2}}$ 
and about 11\% in $\alpha$ had to be assigned.

The dependence on velocity and CO abundance is generally weaker compared to the fits in the literature.
The dependence on the excitation temperature is small.
Of most interest is the dependence on the strength of the ISRF.
Globally, a factor of 15 increase in the ISRF leads to a photodissociation radius that is three times smaller.


\subsection{The effect in two cases} 

As an illustration, the effect of the new photodissociation radii on the CO line intensities is considered in two cases: 
the peculiar OH/IR star OH 32.8$-$0.3, and the star V1 in the cluster 47~Tucanae, which were fitted using the model
in Groenewegen (1994a). It is not the aim here to re-derive the MLRs, but only to show the typical effect a change in photodissociation 
radius may have. A full modelling of an individual star would also consider the line shape of the lines, but here 
only integrated line intensities are compared.

Table~\ref{Tab-Mod} lists the default stellar parameters adopted in Groenewegen (1994b) and McDonald et al. (2015) 
for OH 32.8 and 47Tuc V1, respectively.
Table~\ref{Tab-CO} lists the observed data in the first row (see below), and then the model results, where the second row
gives the result for the photodissociation radii based on MGH.

\begin{table*}

\caption{Sample of stars}
  \begin{tabular}{lccccccccccc}
  \hline
object         & D     & \mdot\    &     V       & $f_{\rm CO}$ & $\Psi$ & $R_{\frac{1}{2}}$ & $\alpha$ &   \\
               & (kpc) & (\msolyr) & (kms$^{-1}$) &        &            &  ($10^{15}$ cm) &          &        \\
\hline 
OH 32.8$-$0.3   & 4.8   & 1.6 (-4) & 15.0 &  6 (-4) & 0.0038   & 1475  & 3.52 & \\ 
47~Tuc V1       & 4.5   & 3.0 (-7) & (10) & 5.3 (-5) & 9.29 (-4) & 12  & 2.11 &  \\  
\hline
\end{tabular}

\label{Tab-Mod}
\end{table*}

\begin{table*}
\setlength{\tabcolsep}{1.4mm}

\caption{CO model results }
  \begin{tabular}{lcccccccl}
  \hline

$\chi$  & $R_{\frac{1}{2}}$ & $\alpha$ &  $T_{\rm int}$ (1-0) &  $T_{\rm int}$ (2-1) &  $T_{\rm int}$ (3-2) & $T_{\rm int}$ (4-3) & $T_{\rm int}$ (6-5) & Remark  \\
        &  ($10^{15}$ cm) &          &     (\kks)         &      (\kks)        &      (\kks)         &       (\kks)     &       (\kks)     & \\
\hline 
        &                &          & \multicolumn{5}{c}{OH 32.8$-$0.3}  \\ \hline
        &                &          & \less 4 & 14.0 $\pm$ 0.5  & 13.1 &  8.5  &  - & observations \\
  -     &   1475.        &   3.52   &    45.3 & 29.2 & 5.23 & 2.66 & 2.68 & $R_{\frac{1}{2}}$ based on MGH \\
  1     &    842.        &   3.04   &    40.8 & 38.7 & 7.29 & 3.04 & 2.85 \\
  1     &    688.        &   2.98   &    30.9 & 37.2 & 7.50 & 3.28 & 2.68 & \mdot = $1 \cdot 10^{-4}$ \msolyr, $\Psi$= 0.0061 \\
  1     &    534.        &   3.06   &    19.1 & 31.1 & 7.29 & 3.87 & 2.72 & \mdot = $5 \cdot 10^{-5}$ \msolyr, $\Psi$= 0.012 \\
  1     &    447.        &   3.15   &    12.6 & 25.8 & 6.98 & 4.36 & 2.90 & \mdot = $3 \cdot 10^{-5}$ \msolyr, $\Psi$= 0.02 \\
  1     &    447.        &   3.15   &    22.5 & 42.9 & 12.5 & 7.97 & 5.44 & \mdot = $3 \cdot 10^{-5}$ \msolyr, $\Psi$= 0.02, $d$= 3.5 kpc \\
  2     &    325.        &   3.39   &    17.8 & 39.4 & 11.8 & 7.91 & 5.51 & \mdot = $3 \cdot 10^{-5}$ \msolyr, $\Psi$= 0.02, $d$= 3.5 kpc \\
 50     &     95.8       &   5.11   &     4.2 & 17.0 &  7.4 & 6.98 & 5.54 & \mdot = $3 \cdot 10^{-5}$ \msolyr, $\Psi$= 0.02, $d$= 3.5 kpc \\
  1     &    447.        &   3.15   &     4.8 & 17.4 &  7.5 & 7.02 & 5.47 & \tablefootmark{a}, $\Psi= 0.02$, $d= 3.5$ kpc \\
  1     &    226.        &   3.97   &     4.6 & 17.4 &  7.5 & 7.02 & 5.47 & \tablefootmark{a}, $\Psi= 0.02$, $d= 3.5$ kpc, $R_{\frac{1}{2}}$ consistent \\
\hline
        &                &          & \multicolumn{5}{c}{47~Tuc V1}  \\ \hline
        &                &          &     -   & \less 0.032  &   -  &   -   & - & observations \\
  -     &   12           &   2.11   &   0.018 & 0.200 & 0.561 & 0.860 & 4.718 & $R_{\frac{1}{2}}$ based on MGH \\
  1     &   19.9         &   2.08   &   0.043 & 0.330 & 0.745 & 1.013 & 4.954 \\
  3     &   10.3         &   2.38   &   0.012 & 0.157 & 0.493 & 0.802 & 4.655 \\
 10     &    5.23        &   2.76   &   0.003 & 0.048 & 0.219 & 0.474 & 3.815 \\
 15     &    4.22        &   2.91   &   0.002 & 0.032 & 0.155 & 0.366 & 3.391 \\ 
 30     &    3.63        &   3.01   &   0.001 & 0.024 & 0.120 & 0.297 & 3.052 \\
 50     &    2.33        &   3.39   &   0.001 & 0.010 & 0.053 & 0.147 & 1.977 \\ 
100     &    1.64        &   3.57   &   0.000 & 0.005 & 0.028 & 0.080 & 1.235 \\ 
 50     &    3.37        &   3.66   &   0.002 & 0.039 & 0.189 & 0.484 & 5.557 &  $\dot{M} \cdot 2$ \\ %
 50     &    2.47        &   3.89   &   0.002 & 0.026 & 0.123 & 0.312 & 3.685 &  $V / 2$                        \\ %
 50     &    2.94        &   3.51   &   0.002 & 0.028 & 0.142 & 0.384 & 4.859 &  $f_{\rm CO} \cdot 2$, $\Psi \cdot 2$  \\  %
\hline
\end{tabular}
\tablefoot{
\tablefoottext{a}{Current MLR of $3 \cdot 10^{-5}$\msolyr. A factor 10 smaller for $r > 1 \cdot 10^{17}$ cm.}
}
\label{Tab-CO}
\end{table*}

\subsubsection{OH 32.8$-$0.3}

For OH 32.8$-$0.3 (V1365 Aql, IRAS 18498-0017) the CO (1-0) and (2-1) data are listed that were taken in the 23 and 11.3\arcsec\ beam 
of the IRAM 30 m telescope and fitted in Groenewegen (1994b).
In addition, more recent data in the J= 3-2 and 4-3 lines are listed for both stars, taken from De Beck et al. (2010). 
The data are taken with the JCMT in beams of 14\arcsec\ and 11\arcsec, respectively. No error bars were provided.

This star comes from the "Group 2" objects defined by Heske et al. (1990). In the objects, the MLR based on 
the CO (1-0) and (2-1) lines is much smaller than that based on the dust, either using the IRAS 60 $\mu$m flux as a tracer, as in  
Heske et al., or from dust RT modelling of the spectral energy distribution (SED), as in Groenewegen 1994b.
This is clear from the first entries in Tab.~\ref{Tab-CO}. The J= 1-0 flux is 10 times stronger then the observed upper limit.
Even if the new photodissociation radius is significantly smaller than the value deduced from MGH the discrepancy remains.

The subsequent models are calculated for a lower MLR, but increasing the dust-to-gas ratio, so that the dust MLR, derived from modelling
the SED, remains constant. The predicted line intensities are lower, but still too large. Also the dust-to-gas ratio becomes 
large at 1/50, i.e. larger than the typically assumed value in the local interstellar medium. 
The next entry is for a model in which the distance is decreased; in fact this model is tuned to fit the observed 
J= 3-2 and 4-3 data. However the J= 1-0 and 2-1 data are overestimated by factors 3-5.

One possible option to decrease the flux is an increased ISRF, which would make the envelope smaller. 
Table~\ref{Tab-CO} lists the results for $\chi = 2$ and the value of 50 that is required to predict J= 1-0 and 2-1 intensities 
in reasonable agreement with observations. 
The J= 3-2 intensity is however now also lower than observed. It is beyond the scope of the paper to assess the likelihood 
of this scenario. However, it would suggest that OH 32.8 is located in or close to a cluster.

An alternative scenario, which was considered in Groenewegen (1994b), is that the MLR is not constant.  
Groenewegen (1994b) considered a MLR a factor 10 lower for radial distances larger than $10^{17}$ cm, corresponding
to a timescale of about 2000 years. Such a model is considered in the last two entries.
In one entry, the photodissociation radius was based on the current MLR; in the second entry, the molecular hydrogen 
density was also lowered for radial distances larger than $10^{17}$ cm and the shielding and resulting CO abundance profile 
was calculated in a consistent way in the photodissociation model. 
In this case the latter improvement does not have an important effect and the J= 1-0 and 2-1 data can be explained by such a model.
In fact the model with a sharp drop in MLR, and the model with a much stronger ISRF predict indistinguishable line intensities.
The models also show that the J= 4-3 and higher transitions are little affected by such extreme models.

\subsubsection{47Tuc V1} 

The model for the variable star V1 in 47~Tuc is unpublished. It was created in connection with McDonald et al. (2015), 
of which MG is a co-author, but the results were not included there; that paper discusses the results from 
two other models. The observed data point is the non-detection of the J= 2-1 line in an ALMA beam of 2\arcsec. 
The model for the 1-0, 3-2, 4-3, and 6-5 data are calculated in beams of 4, 1.3, 1, and 0.33\arcsec.
An expansion velocity of 10~\ks\ is adopted.

The models in Tab.~\ref{Tab-CO} show the result for increasing strength of the ISRF.
For a standard ISRF one would have expected 7-10~$\sigma$ detections, which was the original premise of performing the ALMA observations.
A value of $\chi = 15$ or more is required to have the predicted J= 2-1 line intensity fall below the observed upper limit.
The calculations in McDonald et al. (2015) predicted a 50\% probability of having $\chi > 49$ and this
implies a 1-2~$\sigma$ detection in the J= 3-2 level at best (at similar noise levels). 
The models suggest that the best chance to detect CO in cluster AGB stars is the J= 6-5 (or higher) transition.

The last three entries illustrate the effect of a factor of two increase in the CO abundance, either by increasing the overall MLR, 
decreasing the expansion velocity, or by increasing $f_{\rm CO}$ and the dust-to-gas ratio, as both were scaled from a 
typical solar value using the overall iron abundance in 47~Tuc. Nominally one would expect the CO intensities to 
increase by a factor of two as well, but the increase is larger than that. This could in part be due to subtle RT effects, 
but certainly could also be due to the larger photodissociation radius; the largest intensities are found for the
largest CO envelopes.
These results also show that observations of a single high-J line may yield a detection and thus the determination 
of the crucial expansion velocity, any more detailed modelling would require the detection of (at least) one additional line.

\section{Summary and conclusions}

A numerical code is presented to calculate the CO abundance profile in an envelope under the influence of the ISRF.
This code follows the methodology of Li et al. (2014, 2016) and uses the shielding functions from Visser et al. (2009).
The main limitation of the model is probably that it assumes a homogeneous outflow and does not take into account clumping.
A clumpy CSE and the deep penetration of UV photons has been proposed as a possible mechanism to explain the presence of warm water
in carbon stars (Decin et al. 2010), although the alternative scenario of pulsation-induced shock chemistry (Cherchneff  2011) may
play a more important role in explaining the observations (Lombaert et al. 2016).

A model grid is calculated covering a large parameter space in MLR, expansion velocity, CO abundance, and strength of the ISRF.
Interpolation in the model grid should be sufficient in most cases to determine the photodissociation radius with sufficient
precision, but the code is available upon request for the most detailed analysis when combined with a 
CO line RT code and/or a dust RT code in modelling individual objects.
In such cases the dust parameters (grain size, specific density, dust-to-gas ratio, and UV extinction) can be set consistent
with the dust modelling of the SED,
the CO abundance profile can be used rather than the two-parameter approximation,
the CO excitation temperature profile can be set (based on a CO RT code), and
more complex density structures can be considered, as illustrated in the case of OH 32.8$-$0.3.

One of the most interesting results is the dependence of the photodissociation radius on the ISRF.
Globally, a factor of 15 increase in the ISRF will lead to a three times smaller photodissociation radius.
The effect has been illustrated for the case of 47~Tuc V1.

\begin{acknowledgements}
This research has been funded in part by the Belgian Science Policy Office under contract BR/143/A2/STARLAB.
I would like to thank Joan Vandekerckhove for preparing Figure~1, and
Xiaohu Li for providing the CO abundance profiles of IK Tau and CW Leo in electronic format.
This paper benefitted from discussions with Li, John Black, and Maryam Saberi during the June 2017 COASTARS meeting in Gothenburg, 
and comments on an earlier version of the paper by Iain McDonald.
I used the Dexter software available at \url{http://dc.zah.uni-heidelberg.de/dexter/ui/ui/info} to extract data from published material.

\end{acknowledgements}

\makeatletter
\renewcommand\@biblabel[1]{}
\makeatother

{}

\begin{appendix}

\section{Numerical details}
\label{App-Num}

Input parameters to the model are
the total mass-loss rate in \msolyr, (constant) expansion velocity in the CSE (V, in \ks), 
number ratio of Helium to Hydrogen ($f_{\rm He}$), 
stellar luminosity, and effective temperature of the central star (which gives the stellar radius).
The CO abundance profile is characterised by the abundance ratio relative to H$_2$ at the inner radius 
($f_{\rm CO}$), and estimates for $R_{\frac{1}{2}}$ and $\alpha$.
The dust parameters are the dust-to-gas ratio, $\Psi$, the dust specific density $\rho_{\rm g}$
and grain size $a_{\rm g}$, and the dust extinction coefficient, $Q_{\rm e}$ at 1000 \AA.

At distance $r$ (in units of $10^{15}$cm) the number density of H$_2$ is 
\begin{equation}
n_{\rm H_2} = 1.51 \cdot 10^{13} \, \dot{M} / {\rm V}  / (1 + 4 f_{\rm He})/ r^2,
\end{equation}
and the number density of CO is 
\begin{equation}
n_{\rm CO} = n_{\rm H_2} \; f_{\rm CO} \, \exp\left( - \ln(2) \left( \frac{r}{R_{\frac{1}{2}}} \right) ^{\alpha} \right).
\end{equation}
The dust opacity is the number density of dust particles times the cross-section ($\pi a^2 Q_{\rm e}$) and is
\begin{equation}
n_{\rm d} \sigma_{\rm d} =  3.76 \cdot 10^{-7} \; \frac{Q_{\rm e}}{a_{\rm g} \; \rho_{\rm g}} \; \dot{M}  \Psi/ {\rm V_{d}} / r^2,
\label{Eq-dust}
\end{equation}
where in the present paper the dust velocity (V$_{\rm d}$) is taken equal to the gas velocity (V).
The dust optical depth that is used in Eq.~\ref{Eq-k} is the integral over radius of this quantity,
\begin{equation}
\tau_{\rm dust} = \int n_{\rm d} \sigma_{\rm d} \; d r.
\label{Eq-tau}
\end{equation}

Less important parameters are the inner radius of the shell where the calculation starts (a few stellar radii), and 
the outer radius which is arbitrarily set to 15 $R_{\frac{1}{2}}$. 
Finally, the CO excitation temperature profile needs to be provided.

\begin{figure}

\centering
\includegraphics[width=0.95\hsize]{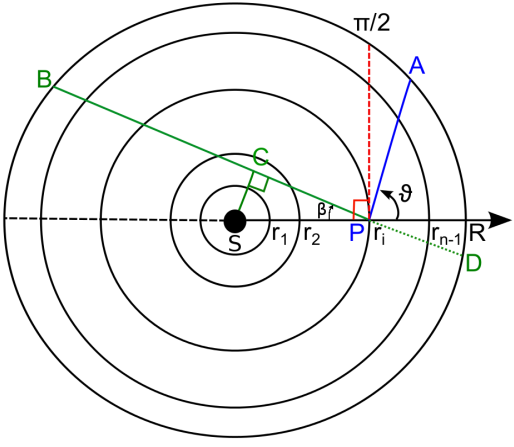}

\caption[]{ 
Structure of the circumstellar envelope model of an AGB star (adopted from Li et al. 2014).
} 

\label{Fig-Sit} 
\end{figure}

The quantities $n_{\rm H_2}$, $n_{\rm CO}$ and $n_{\rm d} \sigma_{\rm d}$ are calculated on a 
radial grid spaced logarithmically between the inner and outer radius.
For $\theta = 0$ (see the structure of the envelope in Fig.~\ref{Fig-Sit}) the column density 
can be determined exactly and compared to the numerical integration. A grid with 200 radial points 
ensures that the column densities are accurate to 0.1\% even for the largest mass-loss rates. 

For each point in the grid $k(r_{\rm i}, \theta)$ (see Eq.~6) is determined which requires the calculation 
of the column densities and optical depth along the lines of sights (los) for all angles $\theta$. 
The procedure in Appendix~A in Li et al. (2014) is followed, which outlines how the case of 
angles $0 < \theta < \pi/2$ (e.g. the los $\overline{PA}$) can be calculated, and how the case of 
   $\pi/2 < \theta < \pi$   (e.g. the los $\overline{PB}$) can be recast into a problem with $\theta = \pi/2$
(the los $\overline{CB} = \overline{CD}$) and the los $\overline{PD}$.

The integral over $\theta$ is split into two parts.
From zero to the angle $(\pi - \beta)$, 
where $\beta$ is the angle where the los $\overline{PB}$ grazes the central star (or the adopted inner 
radius of the envelope), i.e. $\overline{SC} = R_{\star}$, and the angle from $(\pi - \beta)$ to $\pi$, 
which covers the angle subtended by the central object.

The shielding function $\Theta_{\rm H_2, CO}$ is interpolated for the calculated CO and H$_2$ column 
densities and average temperature. The dependence of the shielding on the excitation temperature is small, 
so that assuming a typical temperature is allowed. 
The value $\Theta_{\rm dust}$ follows from $\exp \, (-\tau_{\rm dust})$.

The shielding functions were compared for different angular grid sizes and in the end 
a grid of 130 uniformly distributed points in $\theta$ between 0 and $(\pi - \beta)$ was adopted.

In this way the photodissociation rate at the radial grid points is determined.
However, the integration of Eq.~2 to determine the CO profile requires care as the spatial scale 
for the change in abundance can change by orders of magnitude throughout the envelope. 
The spatial scale for a significant change in abundance at distance $r$ is (V$/I(r)$).
By performing some numerical tests the following procedure was adopted.
If ($f$ V$/I(r)$) (with $f = 0.003$) is smaller than the distance between consecutive grid points, then
the integral in Eq.~2 is evaluated with the photodissociation rate that was previously calculated.
If not, ($f$ V$/I(r)$) is used as stepsize and the photodissociation rate at 
$r_{\rm i} = r_{\rm i-1} + (f$ V$/I(r_{\rm i-1}))$ is determined.
The integration is continued until $x(r_{\rm i}) = 0.001$.

With the profile $x(r)$ calculated, values for $R_{\frac{1}{2}}$ and $\alpha$ are determined by fitting
a straight line to $\ln (- (\ln (x(r)) / \ln(2) ))$ versus $\ln (r)$  over the 
range $0.29 < x(r) < 0.89$.

\begin{figure} 

\centering
\includegraphics[width=0.95\hsize]{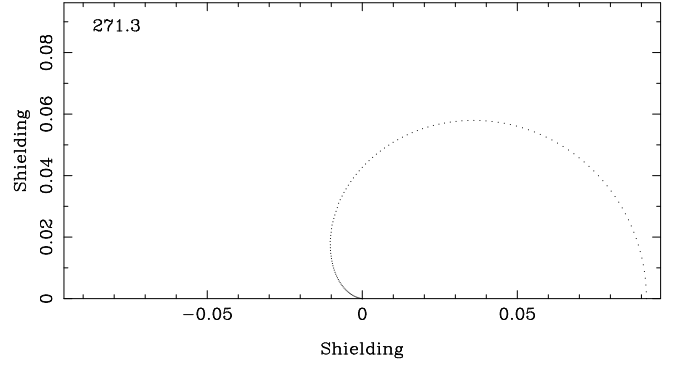}

\caption[]{ 
Total shielding $k (r, \theta)$ as a function of angle near $r \approx R_{\frac{1}{2}}$ for the standard model.
Referring to Eq.~\ref{Eq-k}, the values along the $x-$ and $y-$axis are proportional to 
$(k (r, \theta) \cdot \cos \theta)$, and $(k (r, \theta) \cdot \sin \theta)$, respectively.
} 

\label{Fig-Shield} 
\end{figure}

Figure~\ref{Fig-Shield} illustrates the effect of the shielding. It obviously depends on the parameters of the 
model (shown is the standard model) and on the location in the envelope (shown is $r \approx R_{\frac{1}{2}}$ for that model) 
but the shape is typical. 
The least shielding is in the forward radial direction. 
Shielding increases as the angle $\theta$ increases (Fig.~\ref{Fig-Sit}), in this case by a factor of $\sim 2$ up to $\theta \sim 90\degr$.
Radiation penetrating from the far side of the envelope is greatly reduced.

\end{appendix}

\end{document}